\documentstyle[12pt]{article}
\input psfig
\begin{document}
\def \mpla#1#2#3{Mod. Phys. Lett. A {\bf#1}, #2 (#3)}
\def \nc#1#2#3{Nuovo Cim. {\bf#1}, #2 (#3)}
\def \np#1#2#3{Nucl. Phys. {\bf#1}, #2 (#3)}
\def \pisma#1#2#3#4{Pis'ma Zh. Eksp. Teor. Fiz. {\bf#1}, #2 (#3) [JETP Lett.
{\bf#1}, #4 (#3)]}
\def \pl#1#2#3{Phys. Lett. {\bf#1}, #2 (#3)}
\def \plb#1#2#3{Phys. Lett. B {\bf#1}, #2 (#3)}
\def \pr#1#2#3{Phys. Rev. {\bf#1}, #2 (#3)}
\def \prd#1#2#3{Phys. Rev. D {\bf#1}, #2 (#3)}
\def \prl#1#2#3{Phys. Rev. Lett. {\bf#1}, #2 (#3)}
\def \prp#1#2#3{Phys. Rep. {\bf#1}, #2 (#3)}
\def \ptp#1#2#3{Prog. Theor. Phys. {\bf#1}, #2 (#3)}
\def \rmp#1#2#3{Rev. Mod. Phys. {\bf#1}, #2 (#3)}
\def \rp#1{~~~~~\ldots\ldots{\rm rp~}{#1}~~~~~}
\def \yaf#1#2#3#4{Yad. Fiz. {\bf#1}, #2 (#3) [Sov. J. Nucl. Phys. {\bf #1},
#4 (#3)]}
\def \zhetf#1#2#3#4#5#6{Zh. Eksp. Teor. Fiz. {\bf #1}, #2 (#3) [Sov. Phys. -
JETP {\bf #4}, #5 (#6)]}
\def \zpc#1#2#3{Zeit. Phys. C {\bf#1}, #2 (#3)}
\def\etal{et al.}
\def \indentt{~~~~}
\def\bra#1{\left\langle #1\right|}
\def\ket#1{\left| #1\right\rangle}
\baselineskip=24pt

\begin{titlepage}
\begin{flushright}
Fermilab Conf-95/288 \\
Sept.~1995 \\
\end{flushright}
\vskip 2.0cm
\centerline{\bf $B_s$ Mixing via $\psi K^*$}
\normalsize

\vskip 1.5cm
\centerline{Patricia McBride~\footnote{McBride@fnal.gov}}
\centerline{\it Fermilab, Batavia, IL 60510}

\centerline{Sheldon Stone~\footnote{Stone@suhep.phy.syr.edu}}
\centerline{\it Physics Dept., Syracuse Univ., Syracuse N.Y. 13244}
\vskip 2.0cm

\centerline{\bf Abstract}\vskip 1.0cm
The decay mode $B_s\to\psi \overline{K}^*$ is suggested as a very good way to
measure the $B_s$ mixing parameter $x_s$. These decays can be gathered using a
$\psi\to\ell^+\ell^-$ trigger. This final state has a well resolved four track
decay
vertex, useful for good time resolution and background rejection.
\vskip 5.3cm

\centerline{Presented at }
\centerline{\it BEAUTY '95 - 3rd International Workshop on
             B-Physics at Hadron Machines }

\end{titlepage}
\newpage
\section{Introduction}
\vskip 2mm
\indentt Measurement of $B_s$ mixing would accurately determine one side of
the so called ``unitarity" triangle, because theoretical uncertainties mostly
cancel
when the ratio of $B_s$ to $B_d$ mixing is used \cite{BsBdrat}.

Time dependent mixing measurements using dileptons at LEP have given precise
values for the mixing parameter $x_d$, and have shown that $x_s>8$ at 90\%
confidence level \cite{Stocchi}.  Standard model expectations are that
$60>x_s>12$ \cite{Albur}. In order to
make these measurements the  decay time is calculated according to:
\begin{equation}
t = {L\over {c\beta\gamma}},
\end{equation}
where $L$ is the decay length, c the speed of light and $\beta\gamma$ is equal
to
the momemtum divided by the mass. The error in $t$ is given by
\begin{equation}
\sigma^2_t=\left(\sigma_{L}\over c\beta\gamma \right)^2 +
\left(t\sigma_{\beta\gamma}\over \beta\gamma \right)^2,
\end{equation}
where the first term arises from the error in decay length and the second
arises
from the error in determining the $B_s$ momentum.

Use of several modes have been suggested for  measuring $x_s$  with a $D_s^+$
in
the  final state. They are listed in Table~\ref{table:Bsbr} along with their
predicted branching ratios \cite{Bsbr}.

\begin{table}[th]
\centering
\caption{Branching Ratios for $B_s\to D_s$ Decays}
\label{table:Bsbr}
\vspace*{2mm}
\begin{tabular}{lcc}\hline\hline
Mode & $B_s$ rate & Product branching fraction \\\hline
$D_s^+\ell^-\overline{\nu}$ & 0.105 & $9\times 10^{-3}$\\
$D_s^+\pi^-$ & $(2.6\pm 0.4)\times 10^{-3}$ & $1.1\times 10^{-4}$\\
$D_s^+\pi^+\pi^-\pi^-$ & $(6.3\pm 2.6)\times 10^{-3}$ & $2.7\times 10^{-4}$\\
\hline\hline
\multicolumn{3}{l}{The observable rate for $D_s^+\to K^+K^-\pi^-$ through
the}\\
\multicolumn{3}{l}{intermeadiate states $\phi\pi^+$ and $K^{*o}K^-$ is taken
as 4.3\%.}\\
\end{tabular}
\end{table}

Unfortunately, there are several problems with using these modes to measure
$x_s$. The semileptonic
decay has a large branching ratio, 21\% for the sum of $\mu$ and electron
modes,
 but the undetected neutrino causes the
determination of the $B_s$ momentum to have a relatively large uncertainty.
This limits the maximum possible $x_s$ reach to about 12.

The $D_s^+\pi^-$ mode has a relatively low branching ratio. Furthermore, the
$B$ decay vertex must be
constructed by first forming a $D_s$ decay vertex and then swimming back this
vector to intersect with the $\pi^-$. There also may be combinatorial
background
problems as $D_s$ production in $B$ decays is substantial, about 12\%. The
background problems could be worse in the higher multiplicity $D_s^+\pi^+\pi^-
\pi^-$ mode.

\vskip 2mm
\section{Use of $B_s\to \psi \overline{K}^*$, $\overline{K}^*\to
K^{\mp}\pi^{\pm}$}
\begin{figure} [htbp]
\vspace{-6.8cm}
\centerline{\psfig{figure=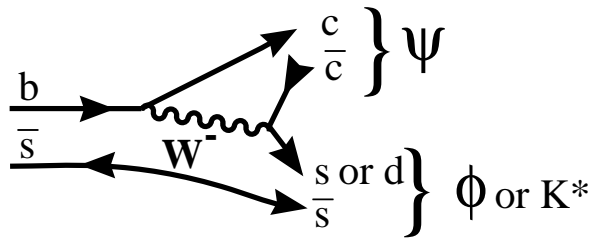,height=5.0in
,bbllx=0bp,bblly=300bp,bburx=600bp,bbury=850bp,clip=}}
\vspace{-.2cm}
\caption{\label{psiks}Weak decay diagrams for $\overline B_s\to \psi\phi$ and
$\psi K^*$. The $K^*$ final state occurs when the virtual $W^-$
materializes as a $\bar{c}d$ pair. }
\end{figure}

\vskip 2mm
\indentt
The simplest $B_s$ final states with $\psi$ mesons are $\psi\eta$ and
$\psi\phi$.
Neither of these can be used to measure $B_s$ mixing because they are not
flavor
specific, i.e. the final state can arise either from a $B_s$ or a
$\overline{B_s}$.
Several years ago the Cabibbo suppressed decay $B_s\to\psi \overline{K}^*$,
$\overline{K}^*\to
K^{\mp}\pi^{\pm}$ was suggested as a possible way to investigate mixing
phenomena \cite{mine}. Here the sign of the kaon charge distinguishes between
$B_s$ or $\overline{B}_s$.This mode would proceed via the diagram shown
in Fig.~\ref{psiks}. The recent
CLEO observation of $B^-\to\psi\pi^-$ decays at the level expected from Cabibbo
suppression is good evidence for the existance of such diagrams \cite{CLpsipi}.
 They measure
\begin{equation}
{{\cal B}(B^-\to\psi\pi^-)\over {\cal B}(B^-\to\psi K^-) }= (5.2\pm 2.6)\%
\approx \lambda^2.
\end{equation}
We predict the branching ratio
\begin{equation}
{\cal B}(B_s\to\psi \overline{K}^*) = {\cal B}(B_d\to\psi K^*)\times \lambda^2=
1.7\times 10^{-3}\times 0.05 = 8.5\times 10^{-5}.
\end{equation}

Another way of describing the yield of these decays is to form
the ratio with respect to $\psi K_s$.  We have
\begin{equation}
{{\cal N}(B_d\to\psi K_s)\over {\cal N}(B_s\to\psi \overline{K}^*)}  = {{\cal
B}(B_d\to\psi
K_s)\times {\cal B}(K_s\to \pi^+\pi^-)\times  {\cal B}(\psi\to\mu^+\mu^-)
\over {\cal B}(B_s\to\psi \overline{K}^*)\times {\cal B}(\overline{K}^*\to
K^-\pi^+)\times  {\cal
B}(\psi\to\mu^+\mu^-)} = 5,
\end{equation}
where ${\cal N}$ indicates the number of observed events, for an equal sample
of $B_d$ and $B_s$.  We have assumed that the detection
efficiency for $\overline{K}^*\to K^-\pi^+$ is equal to that for $K_s \to
\pi^+\pi^-$.  We
expect, however, that the $K_s$ efficiency is significantly lower due to the
long
decay distance of the $K_s$.  We need
to correct for the difference in the relative production ratio between
$B_d$ and $B_s$.    An estimate is obtained
from LEP data by using the measurement of the ratio
of opposite sign dileptons to like sign dileptons and comparing with the same
number found at the $\Upsilon (4S)$, where $B_s$ aren't produced. Such a
calculation gives 1/3-1/4 the number of $B_s$ relative to the number of $B_d$.
Therefore we expect about 1/15 the number of reconstructed and flavor tagged
$B_s\to\psi \overline{K}^*$ as $B_d\to \psi K_s$.
For hadron collider experiments several thousand tagged $\psi K_s$ events
implies
several hundred tagged $\psi \overline{K}^*$ events.

There are several significant advantages using the $\psi \overline{K}^*$ decay
mode.
 A $\psi\to \ell^+\ell^-$ trigger can be used to select these events.
 Furthermore, this decay mode
has a particularly useful topology, having four charged tracks emanating from a
single decay vertex. This is important both for background reduction and for
exquisite decay time resolution.

\begin{figure} [htbp]
\vspace{-2cm}
\centerline{
\psfig{bbllx=70pt,bblly=70pt,bburx=600pt,bbury=600pt,file=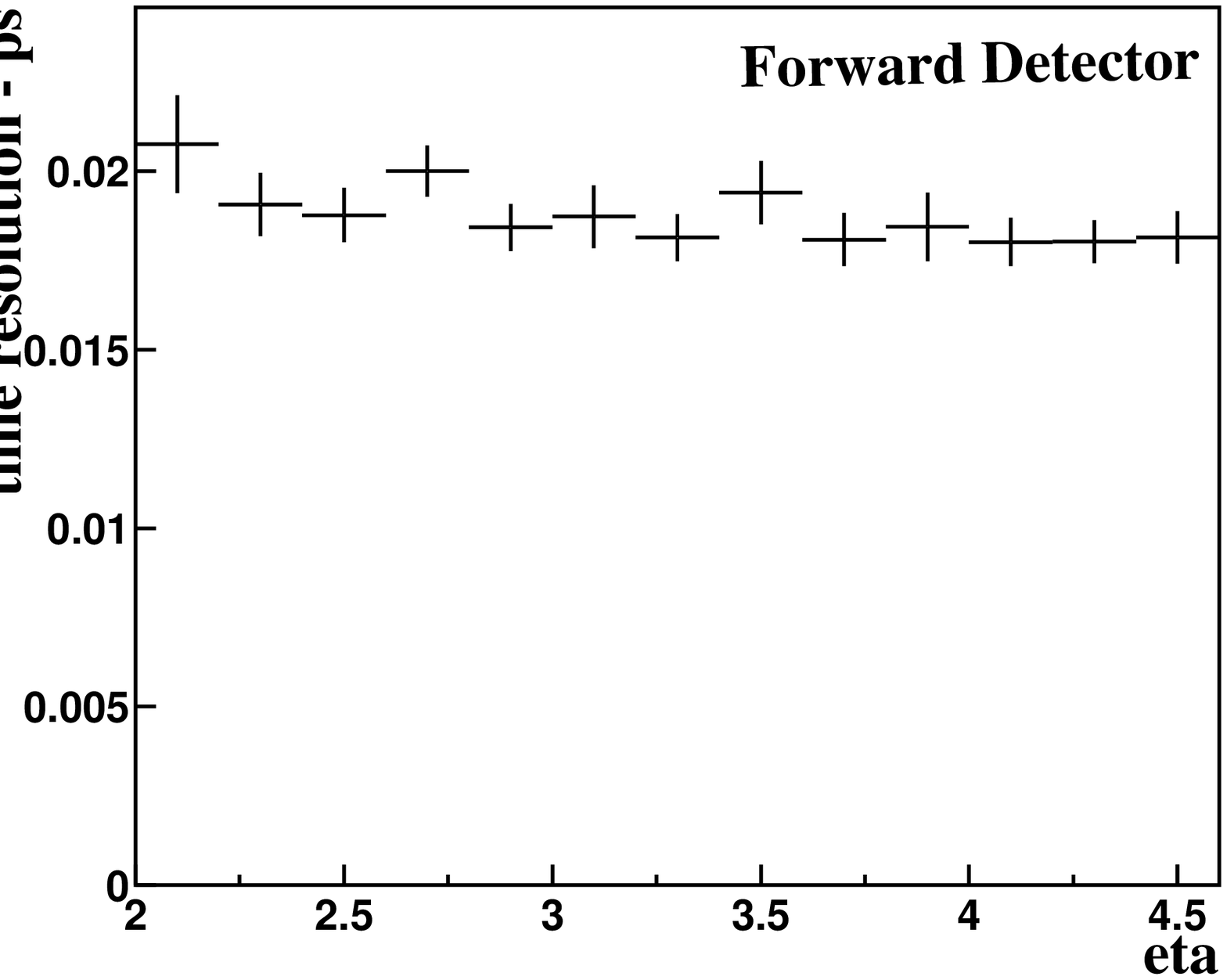,width=2.75in}
\psfig{bbllx=70pt,bblly=70pt,bburx=600pt,bbury=600pt,file=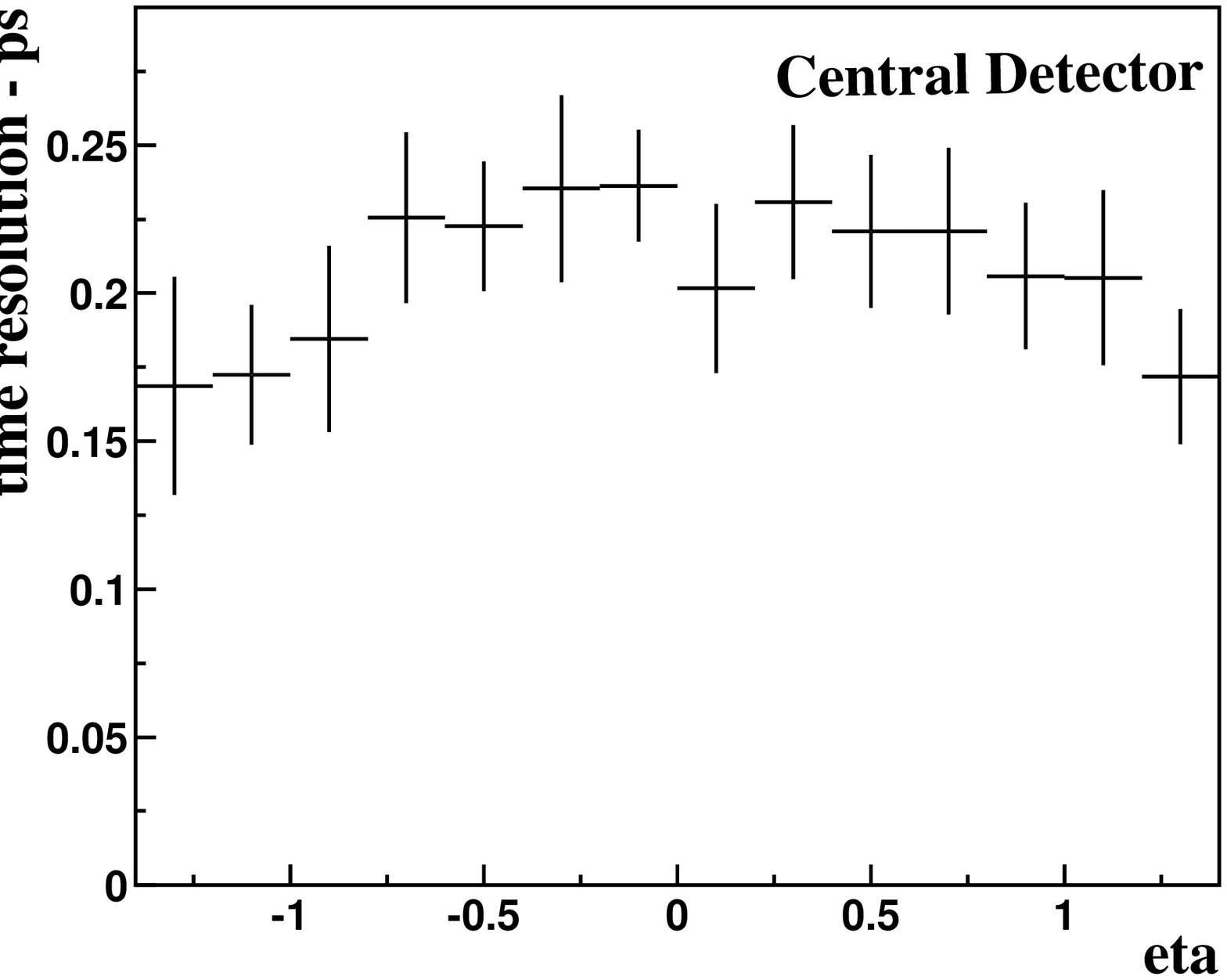,width=2.75in}
}
\vspace{0.5cm}
\caption{\label{dteta}The time resolution plotted as a function of
$\eta$ for a forward detector $(2.0 < \eta < 4.5)$ and a
central detector $(|\eta | < 1.5)$ for the decay $B_s\to\psi \overline{K}^*$
produced at a hadron collider with a center of mass energy of 1.8~TeV.}
\end{figure}

\begin{figure} [htbp]
\vspace{-2cm}
\centerline{
\psfig{bbllx=70pt,bblly=70pt,bburx=600pt,bbury=600pt,file=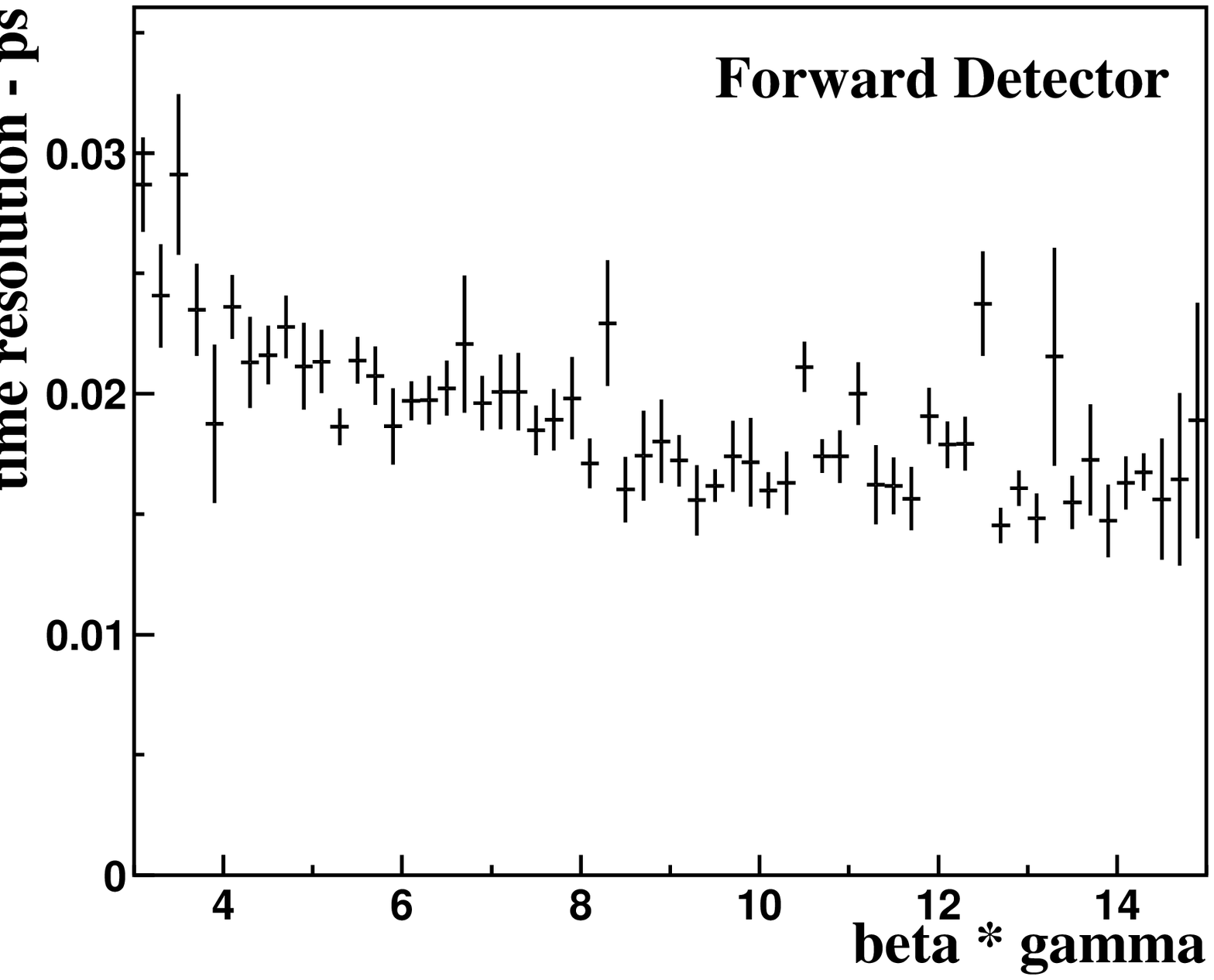,width=2.75in}
\psfig{bbllx=70pt,bblly=70pt,bburx=600pt,bbury=600pt,file=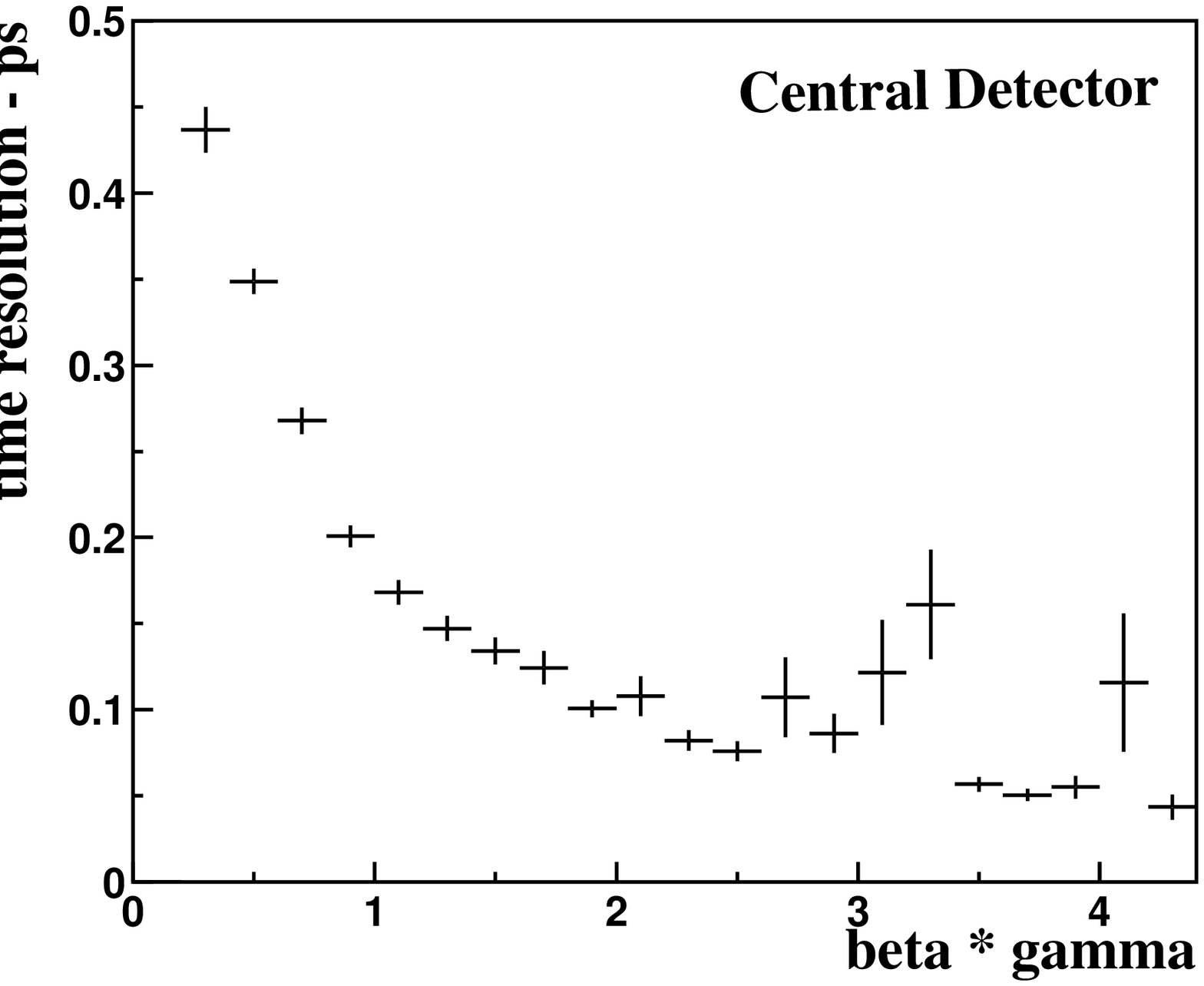,width=2.75in}
}
\vspace{0.5cm}
\caption{\label{dtbg}The time resolution plotted as a function of
$\beta \gamma$ for a forward detector $( 2.0 < \eta < 4.5)$ and a
central detector $(|\eta | < 1.5)$ for the decay $B_s\to\psi \overline{K}^*$
produced at a hadron collider with a center of mass energy of 1.8~TeV.}
\end{figure}

Let us consider the $x_s$ sensitivity. With this fully reconstructed mode the
momentum resolution can be made very good, so there is little effect from the
error in $\gamma$ (second term in equation 2).  We have made estimates of
the time resolution possible in both ``forward" and ``central" detectors at the
FNAL collider. These detectors consist of silicon strips, tracking chambers and
have
a dipole field for the foward detector and a solenoidal field for the central
detector. The simulation program is capable of correctly taking into account
the
track smearing due to multiple scattering and detector resolution, although the
full pattern recognition is not attempted. While detailed resolutions are
subject to
exact detector configurations some clear conclusions have emerged. We show the
time resolution as a function of pseudorapidity ($\eta$) for the forward and
central geometries in Fig. \ref{dteta}.  Similarly the time resolution is
plotted
as a function of $\beta\gamma$ in Fig \ref{dtbg}.

The  time resolution, $\sigma_t$ for the forward geometry is approximately 0.02
ps, while it is
about a factor of 10 worse, 0.2 ps for the central geometry. $\sigma_t$ appears
to be
independent of $\eta$ and independent of $\gamma$ for $\gamma >2.$
 The latter can be understood as being a
result of poorer decay length resolution in direct proportion to $\gamma$, a
well
know effect explained by the folding in spatial angle of the decay products as
$\gamma$ increases. The superiority of the forward geometry results from the
decay tracks not being limited from multiple scattering and the ability to
place
detectors with inherent 10$\mu$m resolution inside the beam pipe.

\begin{figure} [htbp]
\vspace{-1.8cm}
\centerline{\psfig{figure=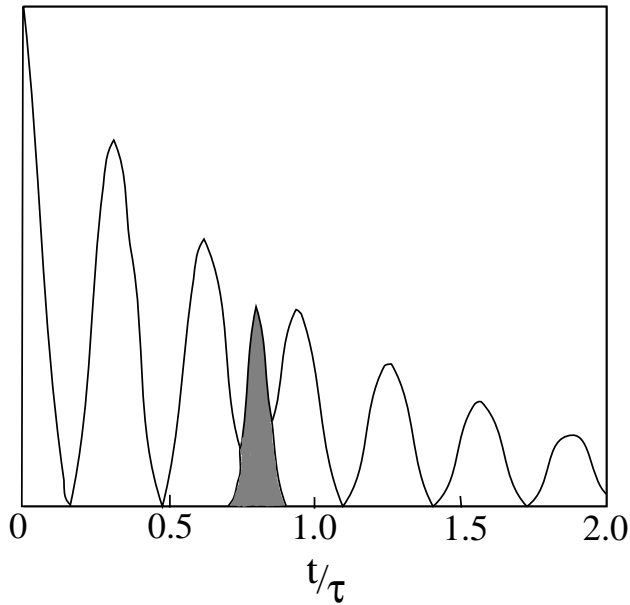,height=5.0in
,bbllx=0bp,bblly=300bp,bburx=600bp,bbury=850bp,clip=}}
\vspace{-2.2cm}
\caption{\label{bsres}The time distribution
$e^-{t\over\tau}\{1+cos\left(x_s{t\over\tau}\right)\}$ for $B_s$ decay
for $x_s=20$. The shaded region shows  a Gaussian with time resolution of
0.05 ps. }
\end{figure}

The number of events as a function of decay time is given by
\begin{equation}
N(t)=N_oe^{-{t\over\tau}}\left(1+cos(x_s{t\over\tau})\right).
\end{equation}
These oscillations are rather rapid on the scale of the $B_s$ lifetime, $\tau$.
A
picture is shown in Fig.~\ref{bsres}, where we have also included a Gaussian
showing the
smearing caused by have a time resolution of 0.05 ps. This resolution was
chosen
from a naive formula, that the time resolution will cause degradation in the
$x_s$
measurement if it is poorer than $1/x_s$ (in units of ps).  Thus, the good time
resolution of the forward FNAL type detector in the $\psi K^*$ mode would allow
a measurement of $x_s$ up to values of approximately 50. The average $\gamma$
for accepted decays is 9.5, giving an average decay length of 4.3 mm. For this
$\gamma$ the resolution in decay length is 50 $\mu$m.

To estimate the real reach in $x_s$ for a particular experimental proposal
requires
studies not only of the vertex resolution, but of the backgrounds and fitting
procedure as well. It is not the purpose of this paper to present such results.
However, the measurement of this channel sets some requirements for a B
detector.
A detector with good vertex resolution will be able to take advantage of the
clean $J/\psi$ signal and the four track
$B$ decay vertex to significantly reduce the background from generic
$B$ decays.  Excellent mass resolution will be needed to eliminate backgrounds
from $B^0_d\to J/\psi K^*$.   Excellent particle identification
will be required to identify the $K$ and $\pi$ in the $K^*$ decay
and to remove background from other channels such as $B_s \to J/\psi\phi$.

Previous
attempts at comparing various experiments \cite{SSCbs} have used a naive
estimate that the
number of tagged $B_s$ required to measure $x_s$ to 20\% of its value (i.e. 5
$\sigma$ requires a number of events:
\begin{equation}
N_{req}= {5^2\over D^2d^2_{time}},
\end{equation}
 where $D$ is the dilution from mistagging including away side mixing and
$d_{time}$ is the dilution from having finite time resolution. $D$ is taken as
approximately 0.5 for most experiments.  For $x_s < 1/\sigma_t$, $d_{time}$ is
close to one. Therefore it appears that it only takes a few hundred fully
reconstructed and tagged $\psi K^*$ events to measure $x_s$ anywhere within
the standard model range. We encourage a full Monte Carlo simulation of this
process.

\section{Conclusions}
\indentt
The decay mode $B_s\to\psi \overline{K}^*$ can be used to measure $x_s$. It is
relatively easy
to trigger on the $\psi\to\ell^+\ell^-$ decay. It has been shown that a forward
detector in a hadron collider has excellent time resolution, of the order of
0.02 ps,
which is sufficient to measure $x_s$ within the standard model range should a
few
hundred tagged events be accumulated.


\begin{thebibliography}{99}
\bibitem{Bsbr}
The $B_s$ branching ratios are assumed to be equal to the $B_d$
branching ratios. In the case of the three pion final state, it is
also assumed that rate for $B_d\to D^{*+}\pi^+\pi^-\pi^-$ is equal
to the rate for $B_d\to D^{+}\pi^+\pi^-\pi^-$.
\end{thebibliography}
\end{document}